\documentclass[prb,aps,10pt]{revtex4}
\begin{document}

\title{Effects of electron-phonon coupling range on the polaron formation}

\author{C. A. Perroni, V. Cataudella, G. De Filippis, V. Marigliano Ramaglia}

\affiliation{Coherentia-INFM  and Dipartimento di Scienze Fisiche, \\
Universit\`{a} degli Studi di Napoli ``Federico II'',\\
Complesso Universitario Monte Sant'Angelo,\\
Via Cintia, I-80126 Napoli, Italy}

\begin {abstract}
The polaron features due to electron-phonon interactions with
different coupling ranges are investigated by adopting a
variational approach. The ground-state energy, the spectral
weight, the average kinetic energy, the mean number of phonons,
and the electron-lattice correlation function are discussed for
the system with coupling to local and nearest neighbor lattice
displacements comparing the results with the long range case. For
large values of the coupling with nearest neighbor sites, most
physical quantities show a strong resemblance with those obtained
for the long range electron-phonon interaction. Moreover, for
intermediate values of interaction strength, the correlation
function between electron and nearest neighbor lattice
displacements is characterized by an upturn as function of the
electron-phonon coupling constant.
\end {abstract}

\maketitle

\newpage

\section {Introduction}

In the last years effects due to strong electron-phonon ($el-ph$)
interactions and polaronic signatures have been evidenced in
several compounds, such as high-temperature cuprate
superconductors, \cite{1}  colossal magnetoresistance manganites,
\cite{2} fullerenes, \cite{fullerene} carbon nanotubes
\cite{carbon} and $DNA$. \cite{3} This amount of experimental data
has stimulated the study of different $el-ph$ coupled systems.
Among the most studied models there are the Holstein lattice model
\cite{4} characterized by a very short-range ($SR$) $el-ph$
interaction, and the Fr$\ddot{o}$hlich model \cite{froh} that
takes into account long-range ($LR$) $el-ph$ couplings in polar
compounds treating the dielectric as a continuum medium. Moreover,
more realistic lattice interaction models including both $SR$ and
$LR$ couplings have been recently introduced. \cite{ahn}

Within the Holstein model a tight-binding electron locally couples
to optical phonon modes. For intermediate $el-ph$ couplings and
electron and phonon energy scales not well separated, it has been
found by numerical studies \cite{7,korn,8,10,ciuchi} and
variational approaches \cite{11,17,17bis} that the system
undergoes a crossover from a weakly dressed electron to a massive
localized polaronic quasiparticle, the small Holstein polaron
($SHP$). A variational approach \cite{17,17bis} proposed by some
of us and based on a linear superposition between the Bloch states
characteristic of the weak and strong coupling regime is able to
describe all the ground state properties with great accuracy.
Moreover this method provides an immediate physical interpretation
of the intermediate regime characterized by the polaron crossover.

Recently a discrete version of the Fr$\ddot{o}$hlich model has
been introduced in order to understand the role of $LR$ coupling
on the formation of the lattice polaron. \cite{alex} Due to the
$LR$ interaction, the polaron is much lighter than the $SHP$ with
the same binding energy in the strong coupling regime.
\cite{alex2} Furthermore the lattice deformation induced by the
electron is spread over many lattice sites giving rise to the
formation of a large polaron ($LP$) also in the strong coupling
region. \cite{fehs} Extending the variational approach previously
proposed for the study of systems with local $el-ph$ coupling,
many properties have been studied by some of us as a function of
the model parameters focusing on the adiabatic regime.
\cite{ourlong} Indeed there is a range of values of the $el-ph$
coupling where the ground state is well described by a particle
with a weakly renormalized mass but a spectral weight much smaller
than unity. Furthermore, with increasing the strength of
interaction in the same regime, the renormalized mass gradually
increases, while the average kinetic energy is not strongly
reduced. Finally due to the $LR$ coupling a strong mixing between
electronic and phononic degrees of freedom has been found even for
small values of the $el-ph$ coupling constant.

Both the $SR$ Holstein and the $LR$ discrete Fr$\ddot{o}$hlich
model can be described by a quite general Hamiltonian $H$

\begin{eqnarray}
H=-t\sum_{<i,j>} c^{\dagger}_{i}c_{j} +\omega_0\sum_{i}
a^{\dagger}_i a_i  +\alpha \omega_0 \sum_{i,j}
f(|\vec{R}_i-\vec{R}_j|) c^{\dagger}_{i}c_{i}\left(
a_j+a^{\dagger}_j\right), \label{1r}
\end{eqnarray}
where $f(|\vec{R}_i-\vec{R}_j| )$ is the interacting force between
an electron on the site $i$ and an ion displacement on the site
$j$. In Eq.(\ref{1r}) $c^{\dagger}_{i}$ ($c_i$) denotes the
electron creation (annihilation) operator at site $i$, whose
position vector is indicated by $\vec{R}_{i}$, and the symbol $<>$
denotes nearest neighbors ($nn$) linked through the transfer
integral $t$. The operator $a^{\dagger}_i$ ($a_i$) represents the
creation (annihilation) operator for phonon on the site $i$,
$\omega_0$ is the frequency of the optical local phonon modes, and
$\alpha$ controls the strength of $el-ph$ coupling. The units are
such that $\hbar=1$. The Hamiltonian (\ref{1r}) reduces to the
Holstein model for
\begin{equation}
f(|\vec{R}_i-\vec{R}_j| )=\delta_{\vec{R}_i,\vec{R}_j},
\label{force1}
\end{equation}
while in the $LR$ case \cite {alex} the interaction
force is given by
\begin{equation}
f(|\vec{R}_i-\vec{R}_j|)=  \left(|\vec{R}_i-\vec{R}_j|^{2} +1
\right)^{-\frac{3}{2}},
\label{force}
\end {equation}
if the distance $|\vec{R}_i-\vec{R}_j|$ is measured in units of
lattice constant.

In addition to the $SR$ and $LR$ case, in this work we analyze the
properties of the system where the electron couples with local and
$nn$ lattice displacements. In this case the interaction force
becomes
\begin{equation} f(|\vec{R}_i-\vec{R}_j|
)=\delta_{\vec{R}_i,\vec{R}_j}+  \frac{\alpha_1}{\alpha}
\sum_{\vec{\delta}} \delta_{\vec{R}_i+{\vec \delta},\vec{R}_j},
\label{force2}
\end{equation}
where ${\vec \delta}$ indicates the $nn$ sites. For all the
couplings of Eqs. (\ref{force1},\ref{force},\ref{force2}) the
$el-ph$ matrix element in the momentum space $M_{\vec{q}}$ is
\begin{equation}
M_{\vec{q}}=\frac{\alpha \omega_0}{\sqrt{L}} \sum_{m}
f(|\vec{R}_m|) e^{i \vec{q} \cdot \vec{R}_m},
\end{equation}
with $L$ number of lattice sites. Through the matrix element
$M_{\vec{q}}$ we can define the polaronic shift $E_p$
\begin{equation}
E_p=\sum_{\vec{q}}\frac{M^{2}_{\vec{q}}}{\omega_0},
\end{equation}
and the coupling constant $\lambda=E_p/zt$, with $z$ lattice
coordination number, that represents a natural measure of the
strength of the $el-ph$ coupling for any range of the interaction.
Limiting the analysis to the one-dimensional case, the matrix
element $M_{\vec{q}}$ is reported in Fig. 1 for the $SR$, $LR$,
and $nn$ extended range ($ER$) couplings. In the $SR$ case the
coupling is constant as function of the transferred phononic
momentum, while in the $LR$ case the vertex is peaked around $q
\simeq 0$. With increasing the ratio $\alpha_1/\alpha$, the $ER$
interaction deviates from the constant behavior developing a peak
around $q \simeq 0$. Actually for the ratio $\alpha_1/\alpha=0.3$
the interaction vertex of the $ER$ case is close to the behavior
of the $LR$ coupling.

In this paper we adopt the variational approach previously
proposed \cite{17,17bis,ourlong} for the study of systems with
local and $LR$ $el-ph$ interactions in order to study the system
with $nn$ $ER$ $el-ph$ coupling. The aim is to investigate the
crossover from $SR$ to $LR$ interactions in the $ER$ model. The
evolution of the ground-state spectral weight, the average kinetic
energy, the mean number of phonons, and the electron-lattice
correlation function with respect to the adiabaticity ratio
$\omega_0/t$ and the $el-ph$ coupling constant is discussed
comparing the results with the local and $LR$ case. For large
values of the $nn$ $el-ph$ coupling many properties show a
behavior similar to those obtained for the $LR$ $el-ph$
interaction. Regions of the model parameters are distinguished
according to the values assumed by the spectral weight giving rise
to a polaronic phase diagram. The transition line between the
crossover and the strong coupling regime continuously evolves
toward that of the $LR$ case by increasing the coupling of the
$ER$ system. Finally, for intermediate values of this coupling,
the correlation function between electron and $nn$ lattice
displacements shows an upturn with increasing the $el-ph$ constant
$\lambda$.

In section II the variational approach is reviewed, while in
section III the results are discussed.

\section {Variational wave function}
In this section the variational approach is briefly summarized.
Details can be found in previous works. \cite{17,17bis}

The trial wave functions are translational invariant Bloch states
obtained by taking a superposition of localized states centered on
different lattice sites
\begin{equation}
|\psi^{(i)}_{\vec{k}}>=\frac{1}{\sqrt{L}}\sum_{\vec{R}_n}e^{i\vec{k}\cdot
\vec{R}_n}|\psi^{(i)}_{\vec{k}}(\vec{R}_n)>,
 \label{12rn}
\end{equation}
where
\begin{equation}
|\psi^{(i)}_{\vec{k}}(\vec{R}_n)> = e^{\sum_{\vec{q}}\left[
h^{(i)}_{\vec{q}}(\vec{k})a_{\vec{q}} e^{i\vec{q}\cdot \vec{R}_n}
+h.c.\right]} \sum_m \phi^{(i)}_{\vec{k}}(\vec{R}_m)
c^{\dagger}_{m+n}|0> .
 \label{13rn}
\end{equation}
In Eq. (\ref{12rn}) the apex $i=w,s$ indicates the weak and strong
coupling polaron wave function, respectively, $|0>$ denotes the
electron and phonon vacuum state, and
$\phi^{(i)}_{\vec{k}}(\vec{R}_m)$ are variational parameters
defining the spatial broadening of the electronic wave function.
The phonon distribution functions $h^{(i)}_{\vec{q}}(\vec{k})$ are
chosen in order to reproduce polaron features in the two
asymptotic limits. \cite{17}

In the intermediate regime the weak and strong coupling wave
functions are not orthogonal and the off-diagonal matrix elements
of the Hamiltonian are not zero. Therefore the ground state
properties are determined by considering as trial state
$|\psi_{\vec{k}}>$ a linear superposition of the weak and strong
coupling wave functions
\begin{equation}
|\psi_{\vec{k}}>=\frac{A_{\vec{k}}
|\overline{\psi}^{(w)}_{\vec{k}}>+ B_{\vec{k}}
|\overline{\psi}^{(s)}_{\vec{k}}>}
{\sqrt{A^2_{\vec{k}}+B^2_{\vec{k}}
+2A_{\vec{k}}B_{\vec{k}}S_{\vec{k}}}},
\label{31r}
\end{equation}
where
\begin{eqnarray}
&&|\overline{\psi}^{(w)}_{\vec{k}}>= \frac{|\psi^{(w)}_{\vec{k}}>}
{\sqrt{<\psi^{(w)}_{\vec{k}}|\psi^{(w)}_{\vec{k}}>}},
|\overline{\psi}^{(s)}_{\vec{k}}>= \frac{|\psi^{(s)}_{\vec{k}}>}
{\sqrt{<\psi^{(s)}_{\vec{k}}|\psi^{(s)}_{\vec{k}}>}} \label{32r}
\end{eqnarray}
and $S_{\vec{k}}$
\begin{equation}
S_{\vec{k}}=
\frac{<\overline{\psi}^{(w)}_{\vec{k}}|\overline{\psi}^{(s)}_{\vec{k}}>+h.c.}
{2}
\label{33r}
\end{equation}
is the overlap factor of the two wave functions
$|\overline{\psi}^{(w)}_{\vec{k}}>$ and
$|\overline{\psi}^{(s)}_{\vec{k}}>$. In Eq.(\ref{31r})
$A_{\vec{k}}$ and $B_{\vec{k}}$ are two additional variational
parameters which provide the relative weight of the weak and
strong coupling solutions for any particular value of $\vec{k}$.
The variational minimization is performed extending the electron
wave function up to fifth neighbors.

\section {Results}

In this section we discuss  ground state properties in the
one-dimensional case for the different ranges of $el-ph$ coupling.

In Fig. 2(a) we report the polaron ground state energy as a
function of the $el-ph$ constant coupling $\lambda$. The
variational method recovers the perturbative results and improves
significantly these asymptotic estimates in the intermediate
region. In this regime the energy decreases with increasing the
range of the $e-ph$ coupling. Moreover, with increasing the range
of the coupling, the crossover between the weak and strong
coupling solution becomes less evident. Actually, as shown in Fig.
2(b), there are marked differences in the ratio $B/A$ that is the
weight of the strong coupling solution with respect to the weak
coupling one. In the $SR$ case the strong coupling solution
provides all the contribution since the overlap with the weak
coupling function is negligible. However, with increasing the
range of the interaction, the weight of the weak coupling function
increases and the polaronic crossover becomes smooth. Another
quantity that gives insight about the properties of the electron
state is the average kinetic energy $K$ reported in Fig. 2(c) (in
units of the bare electron energy). While in the $SR$ case $K$ is
strongly reduced, in the $LR$ case it is only weakly renormalized
stressing that the self-trapping of the electron occurs for larger
couplings with increasing the range of the interaction. Finally
the mean number of phonons is plotted in Fig. 2(d). In the
weak-coupling regime the interaction of the electron with
displacements on different sites is able to excite more phonons.
However, in the strong coupling regime there is an inversion in
the roles played by $SR$ and $ER$ interaction. Indeed the $SHP$ is
strongly localized on the site allowing a larger number of local
phonons to be excited.

In addition to the quantities discussed in Fig. 2, other
properties change remarkably with increasing the ratio
$\alpha_1/\alpha$. An interesting property is the ground state
spectral weight $Z$, that measures the fraction of the bare
electron state in the polaronic trial wave function. As plotted in
Fig. 3(a), the increase of the $el-ph$ coupling strength induces a
decrease of the spectral weight that is more evident with
increasing the range of the $el-ph$ coupling. Only in the strong
coupling regime the spectral weights calculated for different
ranges assume similar small values. While for the local Holstein
model $Z=m/m^*$, as the $ER$ case is considered, $Z$ becomes
progressively smaller than $m/m^*$ in analogy with the behavior
due to the $LR$ interaction. \cite{ourlong} We have found that for
the ratio $\alpha_1/\alpha=0.3$ there is a region of intermediate
values of $\lambda$ where the ground state is described by a
particle with a weakly renormalized mass but a spectral weight $Z$
much smaller than unity. In Fig. 3(b) we propose a phase diagram
based on the values assumed by the spectral weight making a
comparison with that obtained in the $SR$ and $LR$ case.
\cite{17bis,ourlong} Analyzing the behavior of $Z$ it is possible
to distinguish different regimes, for example the crossover regime
($0.1 < Z < 0.9$) characterized by intermediate values of spectral
weight and a mass not strongly enhanced for the $ER$ case, and
strong coupling regime ($Z<0.1$) where the spectral weight is
negligible and the mass is large but not enormous if the range of
the coupling increases. With increasing the range of the
interactions in the adiabatic case there is strong mixing of
electronic and phononic degrees of freedom for values of $\lambda$
smaller than those characteristic of local Holstein interaction.
Furthermore, entering the strong coupling regime, the charge
carrier is still mobile and it does not undergo any abrupt
localization. Only in the antiadiabatic regime the transition
lines separating the crossover from the strong coupling regime
tend to superimpose.

Another important quantity associated to the polaron formation is
the correlation function $S(R_l)$
\begin{equation}
S(R_l)=S_{k=0}(R_l)= \frac{\sum_{n} <\psi_{k=0}
|c^{\dagger}_nc_n\left(a^{\dagger}_{n+l}+a_{n+l}\right)|\psi_{k=0}>}
{<\psi_{k=0} |\psi_{k=0}>} \label{102r}.
\end{equation}
In Fig. 4(a) we report the correlation function $S(R_l=0)$ at
$\omega_0/t=1$ for several ranges of the $el-ph$ interaction. In
analogy with the behavior of the average number of phonons
discussed in Fig. 2(d), the on-site correlation function is larger
with increasing the range of the interaction in the weak coupling
regime, but it becomes smaller as function of the coupling
constant $\lambda$ in the strong coupling region indicating the
the $SHP$ is more effective in producing local lattice
distortions. Actually in the $ER$ and $LR$ case, the lattice
deformation is spread over $nn$ or many lattice sites,
respectively, giving rise to the formation of $LP$ also in the
strong coupling regime. Therefore it is interesting to analyze the
behavior of the correlation function at $nn$ sites. As reported in
Fig. 4(b), in the $SR$ case there is a minimum as function of
$\lambda$ since the particle tends to localize on a single site
with increasing the $el-ph$ coupling. In the $LR$ case the lattice
distortion shows a decreasing behavior with increasing $\lambda$
indicating that the nearest neighbor contribution is always
relevant. However for intermediate values of the ratio
$\alpha_1/\alpha$ in the $ER$ case, the correlation function shows
an upturn as function of the coupling constant $\lambda$.
Actually, for small values of the coupling, this function tends to
follow the behavior of the local interaction, but, with increasing
the value of $\lambda$, the coupling to the $nn$ lattices is able
to give deviations from the $SR$ case. In fact the lattice
deformation reaches a maximum, then begins to decrease following
the behavior of the $LR$ interaction. Therefore, as function of
$\lambda$, two different regimes in the correlation function can
be evidenced.

\section{Discussion and Conclusion}
In this paper we have extended a variational approach in order to
study the polaronic ground-state features of a one dimensional
$el-ph$ model with coupling to local and $nn$ lattice
displacements. Many physical quantities such as the ground state
energy and spectral weight, the average kinetic energy, the mean
number of phonons, and the electron-lattice correlation function
have been discussed making a comparison with the results obtained
with $SR$ and $LR$ interactions. It has been possible to ascertain
that most physical quantities are quantitatively equal to those
obtained for the $LR$ interaction as the $el-ph$ coupling in the
$ER$ case is large. A polaronic phase diagram based on the values
assumed by the spectral weight has been proposed. It has been
shown that the transition lines between the crossover and the
strong coupling regime continuously evolve toward that of the $LR$
case by increasing the coupling of the $ER$ system. The deviations
of the $ER$ case from $LR$ case become evident only in quantities
depending on distances larger than the lattice parameter, such as
in the electron-lattice correlation function. At neighbor nearest
sites for large values of the coupling, the $ER$ interaction is
able to reproduce the correlation function characteristic of the
$LR$, while, at intermediate values of the ratio
$\alpha_1/\alpha$, the lattice deformation shows an upturn as
function of the coupling constant $\lambda$.

Recently, a variational wave function \cite{perrossh} has been
proposed to study the polaron formation in Su-Schrieffer-Heeger
($SSH$) model where the electronic transfer integral depends on
the relative displacement between $nn$ sites. Unlike the original
$SSH$ model, the non-local electron-lattice coupling has been
assumed to be due to the interaction with optical phonon modes. It
has been shown that with this type of interaction the tendency
towards localization is hindered from the pathological sign change
of the effective next-nearest-neighbor hopping. Therefore it is
not possible to reach the strong coupling regime where most
properties obtained with the $ER$ density-type $el-ph$ coupling
bear strong resemblance with those in the $LR$ model. Only the
coupling with acoustic phonons is able to provide a solution with
localized behavior within the $SSH$ model. \cite{lamagna}

The variational approach for models with density-type $el-ph$
coupling can be generalized to high dimensions, where it can still
give a good description of ground state features. \cite{17,giulio}
However, in order to reproduce with the $ER$ interaction most
physical quantities of the $LR$ case, with increasing the
dimensionality, it is important to include not only coupling terms
at $nn$ sites but also at next nearest neighbors. Actually it is
necessary that the expansion of the coupling to near sites gives
rise to an $el-ph$ interaction vertex similar to that obtained in
the $LR$ case. Under these conditions the variational method is
able to interpolate between the behavior of the $SR$ case to the
$LR$ one with increasing the coupling of the interaction with
close sites.

\section*{Figure captions}
\begin {description}

\item{Fig.1}
The $el-ph$ matrix element $M_q$ (in units of $\alpha
\omega_0/{\sqrt L}$) for different ranges of the interaction as
function of the momentum q (in units of $\pi$).

\item{Fig.2}
The ground state energy $E_0$ in units of $\omega_0$ (a), the
ratio B/A at k=0 (b), the average kinetic energy $K$ in units of
the bare one (c) and the average phonon number $N$ (d) for
$t=\omega_0$ as a function of the coupling constant $\lambda$ for
different ranges of the $el-ph$ interaction: $SR$ (solid line),
$ER$ with $\alpha_1/\alpha=0.05$ (dash line), $ER$ with
$\alpha_1/\alpha=0.1$ (dot line),  $ER$ with $\alpha_1/\alpha=0.2$
(dash-dot line), $ER$ with $\alpha_1/\alpha=0.3$ (dash-double dot
line), $LR$ (double dash-dot line).

\item{Fig.3}
(a) The ground state spectral weight at $\omega_0 /t =1$ as a
function of the coupling constant $\lambda$ for different ranges
of interaction: $SR$ (solid line), $ER$ with
$\alpha_1/\alpha=0.05$ (dash line), $ER$ with
$\alpha_1/\alpha=0.1$ (dot line),  $ER$ with $\alpha_1/\alpha=0.2$
(dash-dot line), $ER$ with $\alpha_1/\alpha=0.3$ (dash-double dot
line), $LR$ (double dash-dot line).

(b) Polaron phase diagram for $SR$ (solid line), $ER$ with
$\alpha_1/\alpha=0.2$ (dash-dot line) , $ER$ with
$\alpha_1/\alpha=0.3$ (dash-double dot line), and $LR$ (double
dash-dot line) $el-ph$ interaction. The transition lines
correspond to model parameters such that the spectral weight
$Z=0.1$.

\item{Fig.4}
The electron-lattice correlation functions $S(R_l=0)$ (a) and
$S(R_l=\delta)$ (b) at $\omega_0 /t =1$ for different ranges of
the $el-ph$ interaction: $SR$ (solid line), $ER$ with
$\alpha_1/\alpha=0.05$ (dash line), $ER$ with
$\alpha_1/\alpha=0.1$ (dot line),  $ER$ with $\alpha_1/\alpha=0.2$
(dash-dot line), $ER$ with $\alpha_1/\alpha=0.3$ (dash-double dot
line), $LR$ (double dash-dot line).

\end {description}

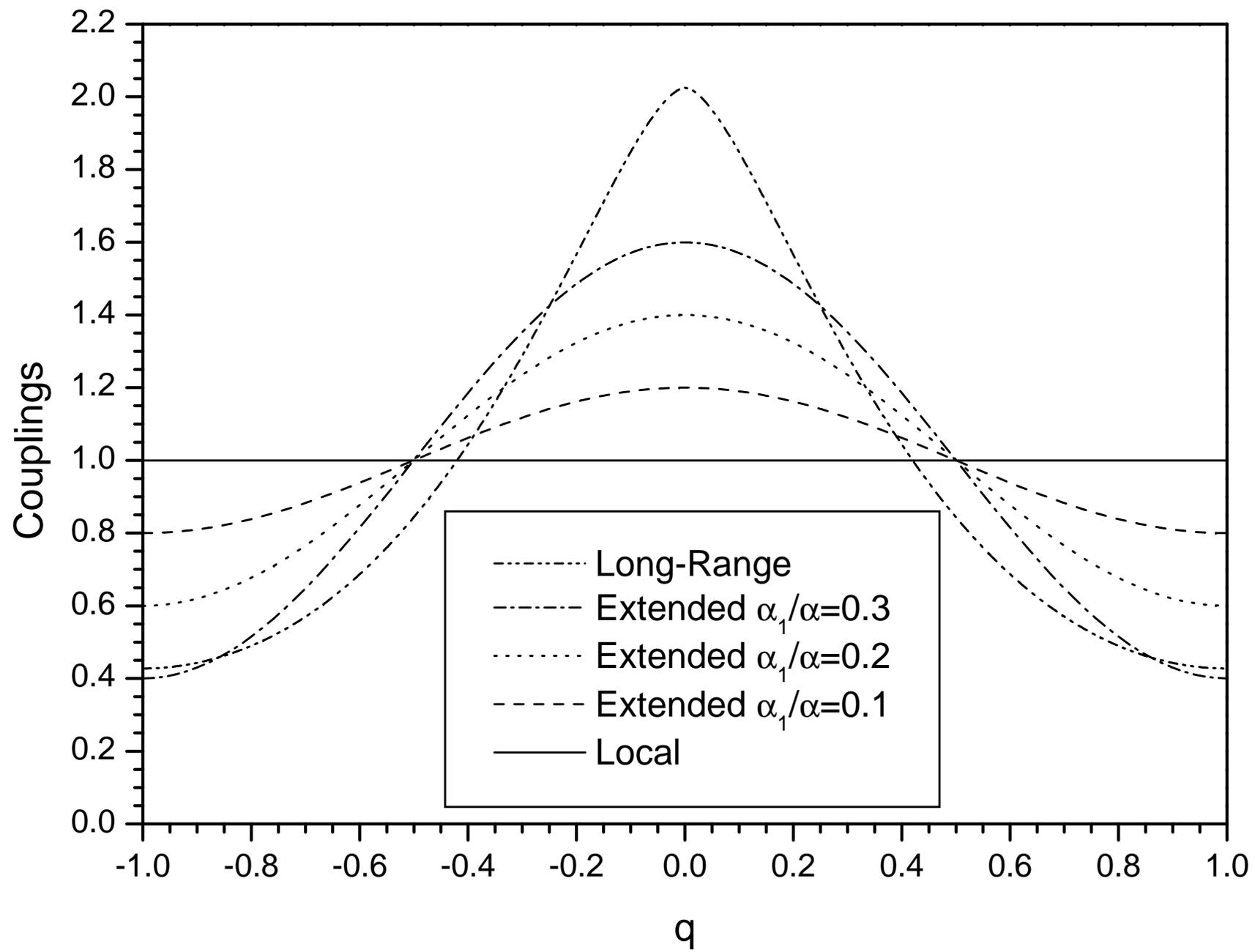
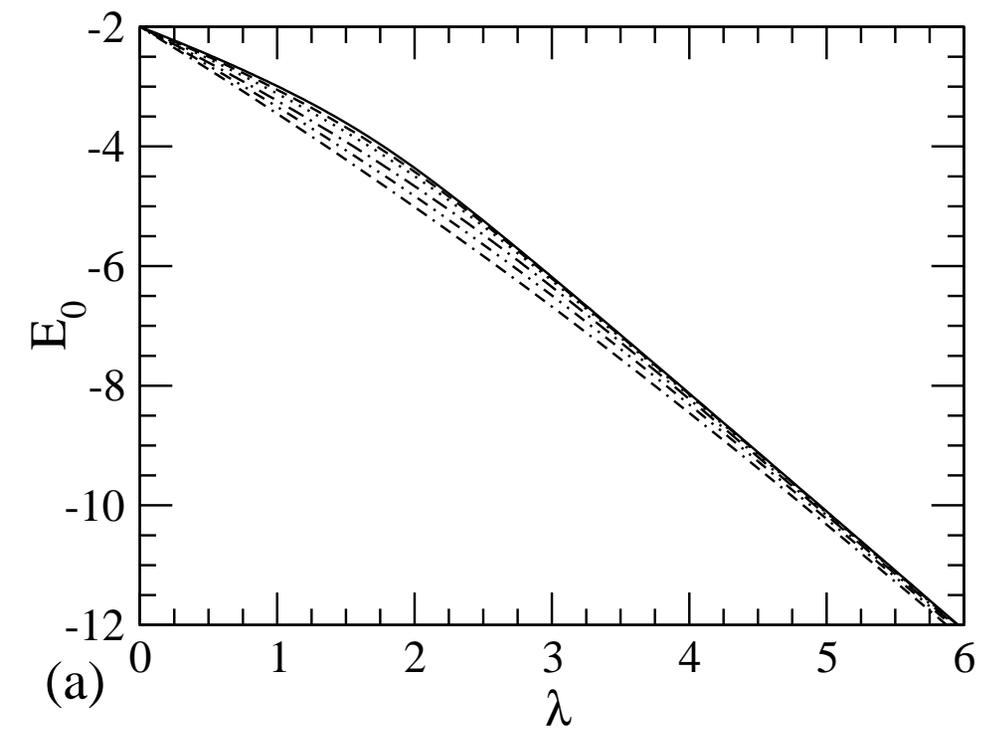
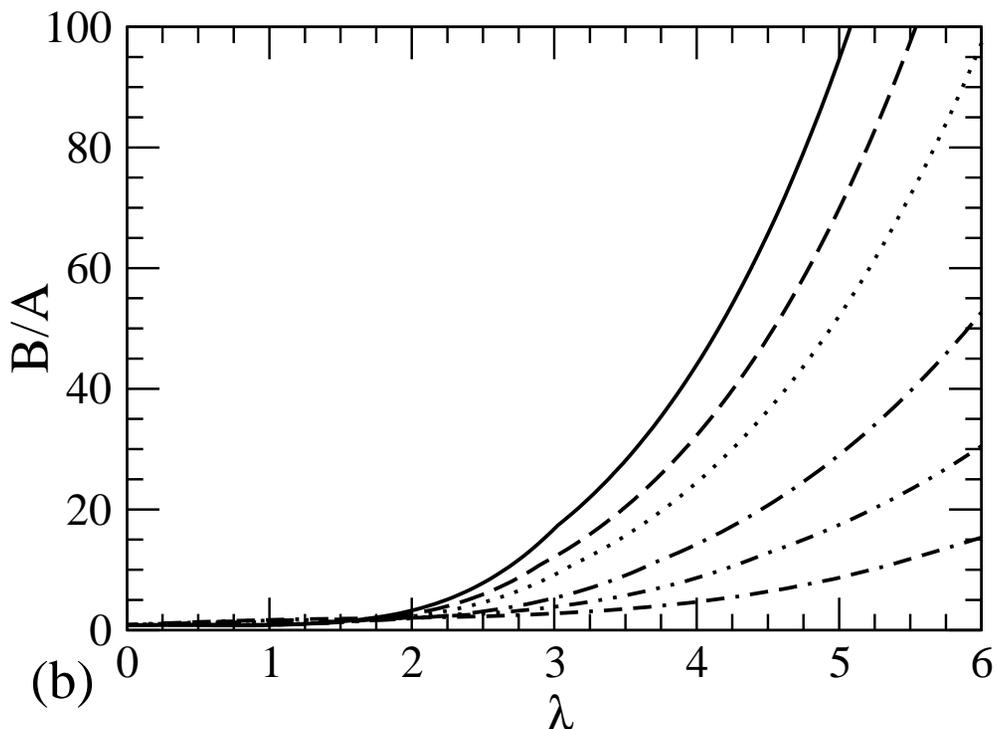
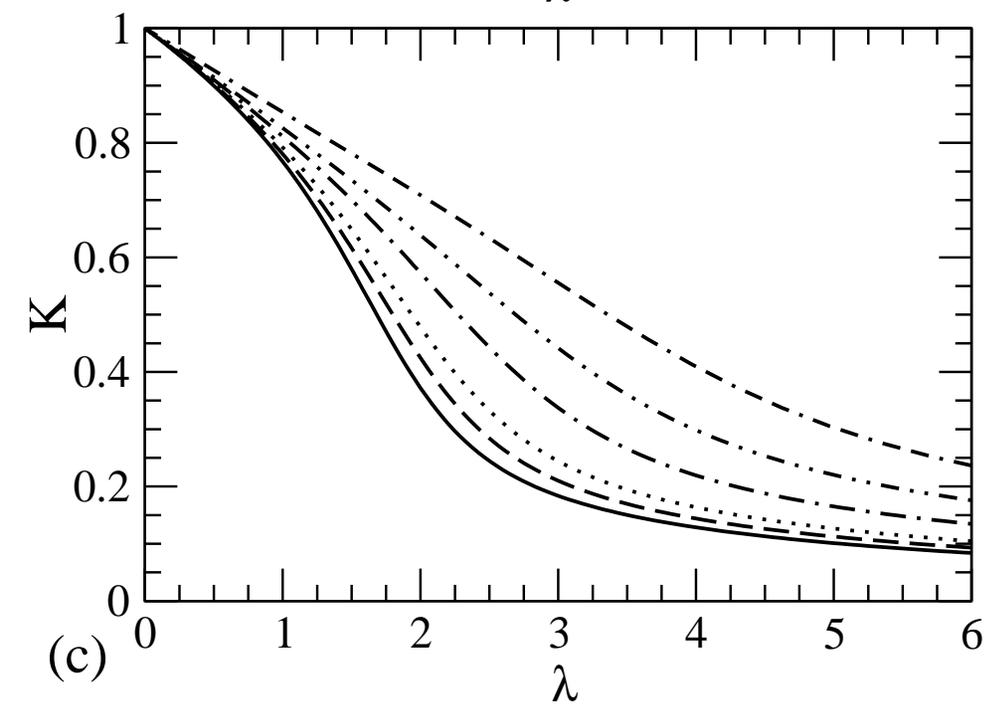
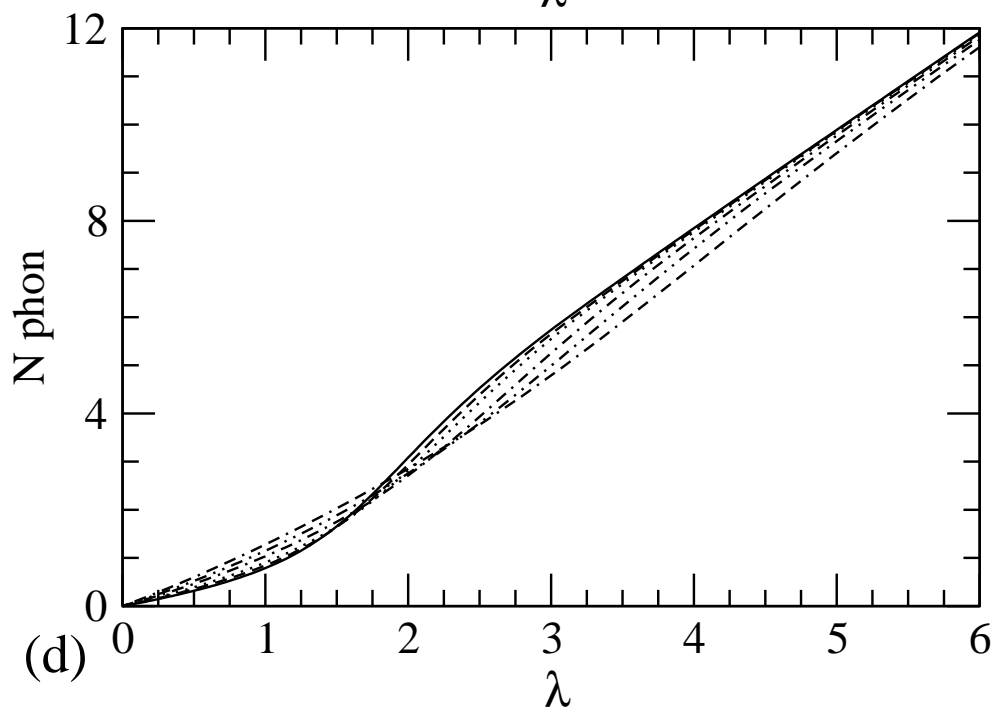
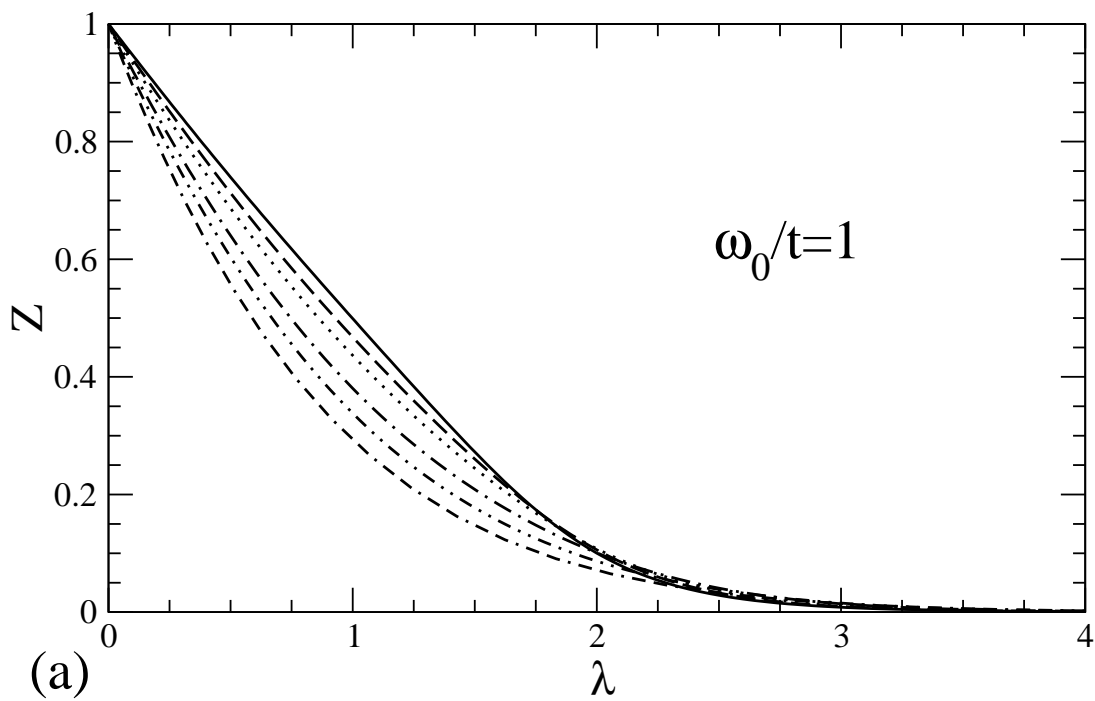
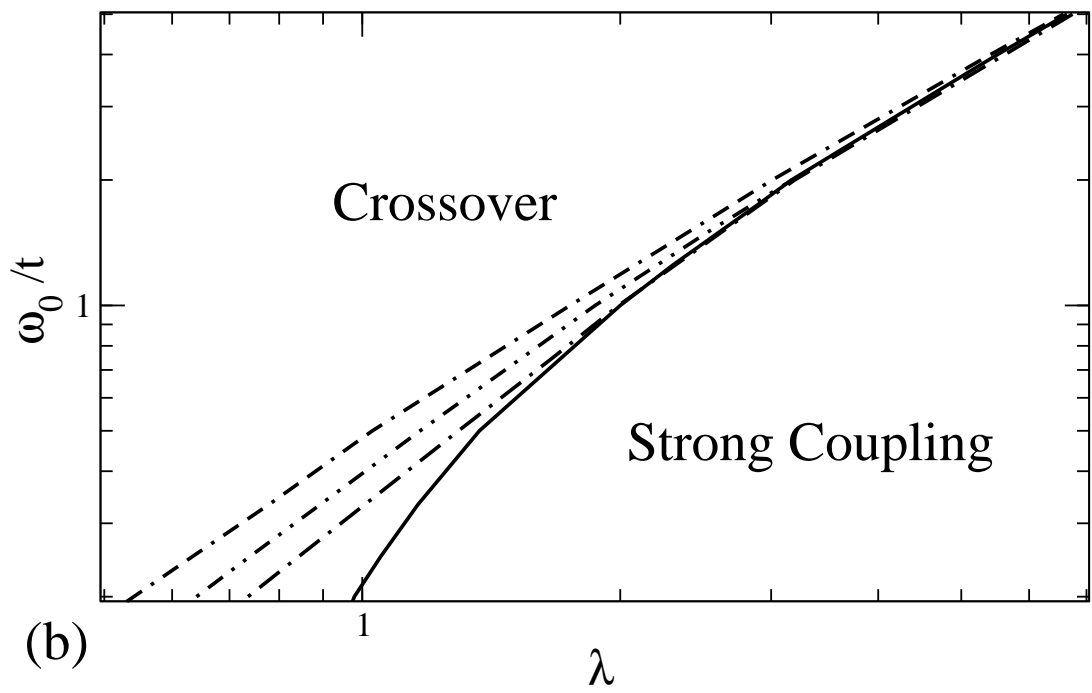
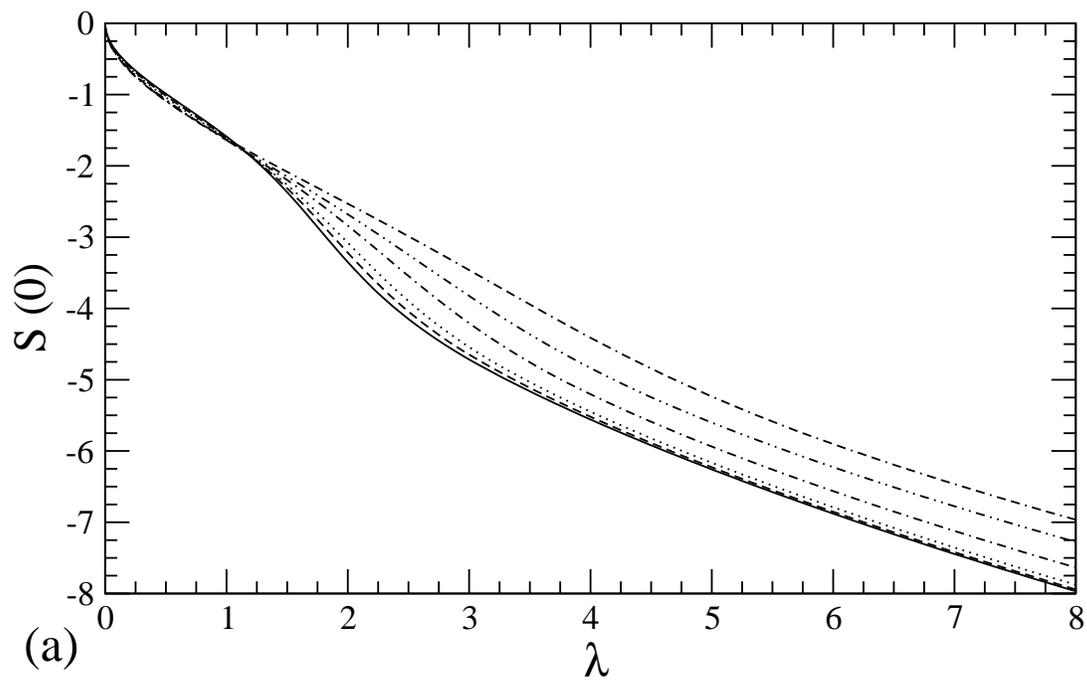
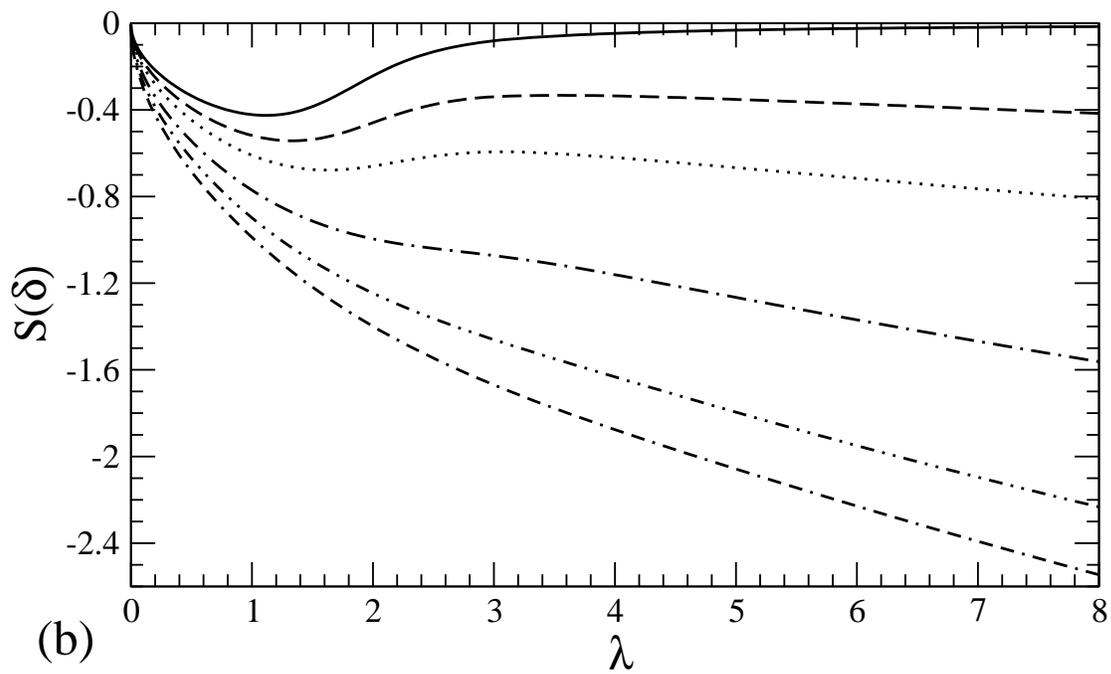
\end{document}